\documentclass[12pt]{article}
\usepackage{amssymb,amsmath,amsfonts,amsthm,amstext,amscd,array}
\usepackage{mathrsfs}
\usepackage{hyperref}
\usepackage{pdfsync}
\usepackage{bbm}
\usepackage{bm}
\usepackage[arrow,matrix,curve]{xy}
\usepackage{bbding}
\usepackage{wasysym}

\usepackage{epsf}
\usepackage{epsfig}
\usepackage{wrapfig}
\def\a{{\sf a}}

\newcommand{\eq}{\begin{equation}}
\newcommand{\eqend}{\end{equation}}
\newcommand{\eqa}{\begin{eqnarray}}
\newcommand{\nonueqa}{\begin{eqnarray*}}
\newcommand{\eqaend}{\end{eqnarray}}
\newcommand{\nonueqaend}{\end{eqnarray*}}

\newcommand{\bma}[1]{\begin{array}{#1}}
\newcommand{\ema}{\end{array}}
\newcommand{\bc}{\begin{center}}
\newcommand{\ec}{\end{center}}

\newcommand{\id}{{1\!\!1}} 

\newif\ifold             \oldtrue

\def\be{\begin{equation}}
\def\ee{\end{equation}}
\def\bea{\begin{eqnarray}}
\def\eea{\end{eqnarray}}
\def\bd{\begin{displaymath}}
\def\ed{\end{displaymath}}

\newcommand{\beq}{\begin{eqnarray}}
\newcommand{\eeq}{\end{eqnarray}}

\makeatletter
\newdimen\normalarrayskip              
\newdimen\minarrayskip                 
\normalarrayskip\baselineskip
\minarrayskip\jot
\newif\ifold             \oldtrue            
\def\arraymode{\ifold\relax\else\displaystyle\fi} 
\def\@arrayskip{\ifold\baselineskip\z@\lineskip\z@
     \else
     \baselineskip\minarrayskip\lineskip2\minarrayskip\fi}
\def\@arrayclassz{\ifcase \@lastchclass \@acolampacol \or
\@ampacol \or \or \or \@addamp \or
   \@acolampacol \or \@firstampfalse \@acol \fi
\edef\@preamble{\@preamble
  \ifcase \@chnum
     \hfil$\relax\arraymode\@sharp$\hfil
     \or $\relax\arraymode\@sharp$\hfil
     \or \hfil$\relax\arraymode\@sharp$\fi}}
\def\@array[#1]#2{\setbox\@arstrutbox=\hbox{\vrule
     height\arraystretch \ht\strutbox
     depth\arraystretch \dp\strutbox
     width\z@}\@mkpream{#2}\edef\@preamble{\halign \noexpand\@halignto
\bgroup \tabskip\z@ \@arstrut \@preamble \tabskip\z@ \cr}%
\let\@startpbox\@@startpbox \let\@endpbox\@@endpbox
  \if #1t\vtop \else \if#1b\vbox \else \vcenter \fi\fi
  \bgroup \let\par\relax
  \let\@sharp##\let\protect\relax
  \@arrayskip\@preamble}
\makeatother

\def\be{\beta}

\newtheorem{definition}[equation]{Definition}

\theoremstyle{definition}

\usepackage{tempora}
\usepackage{color,xcolor}
\usepackage{manfnt}

\def\ddo{\end{document}}
\usepackage{hyperref}
\hypersetup{colorlinks,%
citecolor=black,%
filecolor=black,%
linkcolor=black,%
urlcolor=black}
\def\a{\alpha}

\newcommand{\nc}{\newcommand}
\nc{\lb}{\llbracket}
\nc{\rb}{\rrbracket}
\nc{\gl}{\llbracket}
\nc{\gr}{\rrbracket}

\begin{document}

\title{\Large\bf  Symplectic realization of generalized Snyder-Poisson algebra}
\author{V.G. Kupriyanov and  E.L.F. de Lima   \\
\\{\it Centro de Matem\'atica, Computa\c{c}\~{a}o e
Cogni\c{c}\~{a}o}\\{\it Universidade Federal do ABC, Santo Andr\'e, SP, 
Brasil} \\
      { \footnotesize  e-mail: \texttt{vladislav.kupriyanov@gmail.com, eduardopcmm@gmail.com }}}
\date{\today} 
\maketitle
\begin{abstract}
We investigate Snyder space-time and its generalizations, including Yang and Snyder–de Sitter spaces, which constitute manifestly Lorentz-invariant noncommutative  geometries. This work initiates a systematic study of gauge theory on such spaces in the semi-classical regime, formulated as Poisson gauge theory. As a first step, we construct the symplectic realizations of the relevant noncommutative spaces, a prerequisite for defining Poisson gauge transformations and field strengths. We present a general method for representing the Snyder algebra and its extensions in terms of canonical phase-space variables, enabling both the reproduction of known representations and the derivation of novel ones. These canonical constructions are employed to obtain explicit symplectic realizations for the Snyder–de Sitter space and to construct the deformed partial derivative which differentiates the underlying Poisson structure. Furthermore, we analyze the motion of freely falling particles in these backgrounds and comment on geometry of the associated spaces.
\end{abstract}

\section{Introduction}
\label{sec:intro}

Noncommutative geometry has long been considered a credible framework for Planck-scale physics~\cite{doplicher1994spacetime}. In particular, Snyder’s 1947 construction introduced a Lorentz-invariant noncommutative spacetime with a fundamental length \cite{snyder_quantized_1947}, preserving Lorentz symmetry while regularizing short-distances behavior. Quantum-gravity and string-theoric arguments further suggest that noncommutativity and uncertainty relations may emerge near the Planck scale \cite{doplicher1994spacetime, lukierski_quantum_1994}, implying minimal length and/or momentum scales \cite{kempf_quantum_1992,maggiore_algebraic_1993} and a lower bound on spatial localization \cite{Padmanabhan:1985jdl, Veneziano:1986zf}.

An elegant realization of Snyder's idea uses a five-dimensional flat space of signature $(1,4)$, with physical spacetime described as a projective slice of the de Sitter(dS) hyperboloid \cite{yang_quantized_1947}
\begin{equation}
y_A\, y^A = R^2,
\qquad
y_A \,y^A := -y_0^2 +y_1^2 + y_2^2 +y_3^2 + y_4^2\,.
\end{equation}
In this picture the Lorentz sector is undeformed, while translations act nonlinearly on coordinates. Shortly after, Yang embedded positions, momenta, and Lorentz generators into an $so(1,5)$ algebra, unifying Snyder’s minimal length with dS constant curvature ~\cite{yang_quantized_1947} and making Born’s reciprocity between positions and momenta manifest ~\cite{born_reciprocity_1949}. The Yang Poisson algebra \cite{Bilac:2024wex} consists of the algebra of Poisson brackets containing the usual Lorentz
algebra of generators $M_{\mu\nu}$ with a standard action on phase space variables ,
\begin{eqnarray}
&&\{M_{\mu\nu},M_{\rho\sigma}\}=\eta_{\mu\rho}\,M_{\nu\sigma}-\eta_{\mu\sigma}\,M_{\nu\rho}-\eta_{\nu\rho}\,M_{\mu\sigma}+\eta_{\nu\sigma}\,M_{\mu\rho}\,,\label{M1}\\
&&\{M_{\mu\nu},x_\rho\}=\eta_{\mu\rho}\,x_\nu-\eta_{\nu\rho}\,x_\mu\,,\qquad \{M_{\mu\nu},p_\rho\}=\eta_{\mu\rho}\,p_\nu-\eta_{\nu\rho}\,p_\mu\,.\label{M2}
\end{eqnarray}
In addition, \emph{both} coordinates and momenta are noncommutative, with Poisson brackets proportional to the Lorentz generators:
\begin{eqnarray}\label{gS}
\{x_\mu,x_\nu\}=\beta^2\,M_{\mu\nu}\,,\qquad \{p_\mu,p_\nu\}=\alpha^2\,M_{\mu\nu}\,.
\end{eqnarray}
To close the algebra it is introduced an additional generator $h$ satisfying,
\begin{equation}\label{h}
\{x_\mu,x_\nu\}=\eta_{\mu\nu}\,h\,,\qquad \{h,x_\mu\}=\beta^2p_\mu\,,\qquad \{h,p_\mu\}=-\alpha^2x_\mu\,,\qquad \{h,M_{\mu\nu}\}=0\,.
\end{equation}

Within the broader Doubly Special Relativity (DSR) \cite{amelino-camelia_testable_2001} program and its three-scale extension, Triply Special Relativity (TSR) \cite{kowalski-glikman_triply_2004}, also known as Snyder–de Sitter (SdS) model provides a manifestly Lorentz-invariant noncommutative spacetime on a constant-curvature background \cite{guo_yangs_2008,PhysRevD.82.105031,mignemi_classical_2012}. The latter is represented in the semiclassical limit by the set of Poisson brackets,
\begin{eqnarray}
\label{SdS}
\{x_\mu,x_\nu\}&=&\beta^2\left(x_\mu\,p_\nu-x_\nu\,p_\mu\right),\qquad \qquad \{p_\mu,p_\nu\}=\alpha^2\left(x_\mu\,p_\nu-x_\nu\,p_\mu\right),\\
\{x_\mu,p_\nu\}&=&\eta_{\mu\nu}+\beta^2p_\mu\,p_\nu+\alpha^2x_\mu x_\nu+2\alpha\beta p_\mu x_\nu\,,\notag
\end{eqnarray}
with the Lorentz generator given by  $M_{\mu\nu}=x_\mu\,p_\nu-x_\nu\,p_\mu$.
Two observer-independent deformation parameters appear: $\beta$, which controls position-space noncommutativity (a length scale), and $\a$, controlling momentum-space curvature (the inverse dS radius). In the limits $\a \to 0$ and  $\beta \to 0$ one recovers, respectively, Snyder algebra \cite{battisti2009modification,Pachol1} ,
\begin{eqnarray}
\label{Sn}
\{x_\mu,x_\nu\}=\beta^2\left(x_\mu\,p_\nu-x_\nu\,p_\mu\right),\qquad \{x_\mu,p_\nu\}=\eta_{\mu\nu}+\beta^2p_\mu\,p_\nu\,,\qquad\{p_\mu,p_\nu\}=0\,,
\end{eqnarray}
 and the Heisenberg algebra on dS space in projective coordinates. SdS (and its anti-dS counterpart) realizes a Born-type exchange symmetry between positions and momenta and exhibits characteristic lower bounds on localization in appropriate sectors \cite{mignemi_classical_2012}. A complementary viewpoint is provided by TSR contractions: for $R \to \infty$ one recovers DSR/$\kappa$-Poincaré–type kinematics in flat spacetime (curved momentum space of scale $\kappa$), while for $\kappa \to \infty$ one obtains the phase-space algebra of a particle on dS spacetime; in both cases a Born-type reciprocity emerges naturally~\cite{guo_yangs_2008}. Further generalizations and $\kappa$-deformations of the Snyder and SdS algebras can be found in \cite{kappa-def,kappa-def1}.

Against this background, we initiate a systematic study of gauge theory on Snyder-type spaces in the semiclassical regime, known as Poisson gauge theory or Poisson electrodynamics \cite{PGT}-\cite{Kurkov:2025abv}. A key mathematical ingredient in this framework is the symplectic realization of the underlying noncommutative structure, which is essential for defining Poisson gauge transformations \cite{KS-21} and the associated field strengths \cite{Kupriyanov:2022ohu}. The main objective of this work is to construct a symplectic realization of the Snyder algebra and its extensions. Since the algebra itself is already symplectic, the most straightforward approach consists in representing it in terms of canonical (Darboux) coordinates on phase space and then extending the phase space in a trivial manner. 

The work is organized as follows. In Section 2 we formulate the systematic approach to the construction of canonical phase space representation of generalized SdS/Yang Poisson algebra. By considering the simplest solutions for the canonical ansatz, parametrized by the functions $a,b,c,d$, we recover known canonical phase-space realizations of the generalized Snyder algebra and identify several new ones. Section 3 studies the flat/commutative regimes and the induced geometry of free fall, extracting effective metrics and commenting on curvature. Section 4 builds the symplectic realization necessary for Poisson gauge theory: extended Darboux variables, the $\Gamma^{{\cal M}}_{{\cal N}}(X)$ matrix, with $X^{\mathcal{M}}= (x^\mu, p_\mu)$, and twisted parcial derivatives $\bar\partial_{\cal N}:=\Gamma^{{\cal M}}_{{\cal N}}(X)\,\partial_{\cal M}$  that restore a Leibniz rule $\bar\partial_{\cal N}\{f,g\}=\{\bar\partial_{\cal N} f, g\}+\{f,\bar\partial_{\cal N} g\}$ with respect to the original Poisson brackets (\ref{SdS}). Conclusions summarize implications for Poisson gauge dynamics and quantization.

\section{Generalized Snyder-Poisson algebra and\\
canonical phase space representations}
\label{sex:algebra}

Consider the generalized Snyder-Poisson algebra in which the commutator of both coordinates and momenta are proportional to the Lorentz generators as stated in (\ref{gS}) with (\ref{M1}) and (\ref{M2}). The mixed bracket takes the general form
\begin{equation}\label{gSg}
\{x_\mu,p_\nu\}=\gamma_{\mu\nu}(x,p)=\eta_{\mu\nu}+\dots\,,
\end{equation}
where the choice of the tensor $\gamma_{\mu\nu}(x,p)$ will define the model (Snyder, SdS, Yang, etc.).
 Our goal is to realize the algebra (\ref{M1}-\ref{gS},\ref{gSg}) on a Darboux (canonical) phase space defined by
\begin{equation}\label{Dc2na}
\{\bar y^\mu,\bar y^\nu\}=0\,,\qquad \{\bar y^\mu,\bar \xi_\nu\}=\delta^\mu_\nu\,,\qquad \{\bar\xi_\mu,\bar\xi_\nu\}=0\,,
\end{equation}
in such a way that we should be able to recover the  realization of the Snyder algebra (\ref{Sn}) in the flat limit $\alpha\to 0$. Then we compute the corresponding $\gamma_{\mu\nu}(x,p)$ which will determine the model.

We start analyzing the conditions \eqref{M1}–\eqref{M2} from which it follows that the Lorentz generators, coordinates and momenta can be represented as
\begin{equation}\label{M0}
M_{\mu\nu}=\bar y_\mu\,\bar\xi_\nu- \bar y_\nu\,\bar\xi_\mu\,,
\end{equation}
and
\begin{equation}\label{M3}
x_\mu=a(u,v,z)\,\bar y_\mu+b(u,v,z)\,\bar\xi_\mu\,,\qquad p_\mu=c(u,v,z)\,\bar y_\mu+d(u,v,z)\,\bar\xi_\mu\,,
\end{equation}
where the Lorentz invariants are $u=1/2\bar y_\mu\bar y^\mu$, $v=1/2\bar \xi_\mu\bar \xi^\mu$ and $z=\bar y_\mu\bar \xi^\mu$. The inverse read,
\begin{equation}\label{M4}
\bar y_\mu=\frac{d\,x_\mu-b\,p_\mu}{ad-bc}\,,\qquad \bar \xi_\mu=\frac{a\,p_\mu-c\,x_\mu}{ad-bc}\,.
\end{equation}
Substituting (\ref{M4}) in (\ref{M0}) one finds,
\begin{equation}
M_{\mu\nu}=\frac{x_\mu\,p_\nu-x_\nu\,p_\mu}{ad-bc}\,.
\end{equation}
To be able to reproduce the standard Snyder algebra (\ref{Sn}) or Snyder-de-Sitter model (\ref{SdS}) in which $M_{\mu\nu}=x_\mu\,p_\nu-x_\nu\,p_\mu$ we shall impose
\begin{equation}
ad-bc=1\,.
\end{equation}
 Let us calculate the Poisson bracket,
\begin{eqnarray}
\{x_\mu,x_\nu\}&=&\{a(u,v,z)\,\bar y_\mu+b(u,v,z)\,\bar\xi_\mu,a(u,v,z)\,\bar y_\nu+b(u,v,z)\,\bar\xi_\nu\}\\
&=&\left[\{a,b\}+\partial_z(ab)-a\,\partial_va-b\,\partial_ub\right]\left(\bar y_\mu\,\bar\xi_\nu- \bar y_\nu\,\bar\xi_\mu\right).\notag
\end{eqnarray}
So, the first of the relations (\ref{gS}) holds if,
\begin{equation}
\{a,b\}+\partial_z(ab)-a\,\partial_va-b\,\partial_ub=\beta^2\,.
\end{equation}
Doing the same with the Poisson bracket of momenta in (\ref{gS}) one finds the equation,
\begin{equation}
\{c,d\}+\partial_z(cd)-c\,\partial_vc-d\,\partial_ud=\alpha^2\,.
\end{equation}
For the reason of completeness let us also calculate,
\begin{eqnarray}\label{pbxp}
\{x_\mu,p_\nu\}&=&\{a(u,v,z)\,\bar y_\mu+b(u,v,z)\,\bar\xi_\mu,c(u,v,z)\,\bar y_\nu+d(u,v,z)\,\bar\xi_\nu\}\\
&=&\left(ad-bc\right)\eta_{\mu\nu}+\bar y_\mu\,\bar y_\nu\left(\{a,c\}+a\,\partial_zc-c\,\partial_za+d\,\partial_ua-b\,\partial_uc\right)+\notag\\
&&\bar y_\mu\,\bar \xi_\nu\left(\{a,d\}+\partial_z(ad)-c\,\partial_va-b\partial_ud\right)+\notag\\
&&\bar \xi_\mu\,\bar y_\nu\left(\{b,c\}+a\,\partial_vc+d\,\partial_ub-\partial_z(bc)\right)+\notag\\
&&\bar \xi_\mu\,\bar \xi_\nu\left(\{b,d\}+d\,\partial_zb-b\,\partial_zd+a\,\partial_vd-c\,\partial_vb\right).\notag
\end{eqnarray}

Taking into account,
\begin{equation}
    \{u,v\}=z, \qquad \{u,z\}=2u, \qquad \{v,z\}=-2v, 
\end{equation}
for the Poisson brackets of the functions $a(u,v,z), b(u,v,z)$ with respect to the canonical bracket on $(\bar{y}, \bar{\xi})$ one finds
\begin{eqnarray}
\{a(u,v,x),b(u,v,z)\}&=&2u\left(\partial_ua\,\partial_zb-\partial_za\,\partial_ub\right)+2v\left(\partial_za\,\partial_vb-\partial_va\,\partial_zb\right)\notag\\
&&+z\left(\partial_ua\,\partial_vb-\partial_va\,\partial_ub\right),
\end{eqnarray}
and similarly for $\{a,c\}$, etc. Altogether, the conditions (\ref{gS}-\ref{M2}) hold true if the Lorentz generators $M_{\mu\nu}$, coordinates $x_\mu$ and momenta $p_\mu$ are given by (\ref{M0}) and (\ref{M3}) correspondingly, where the functions $a$, $b$, $c$ and $d$ should satisfy the equations,
\begin{eqnarray}
&&2u\left(\partial_ua\,\partial_zb-\partial_za\,\partial_ub\right)+2v\left(\partial_za\,\partial_vb-\partial_va\,\partial_zb\right)\notag+z\left(\partial_ua\,\partial_vb-\partial_va\,\partial_ub\right)\notag \\
&&+\partial_z(ab)-a\,\partial_va-b\,\partial_ub=\beta^2\,,\label{s1}\\
&&2u\left(\partial_uc\,\partial_zd-\partial_zc\,\partial_ud\right)+2v\left(\partial_zc\,\partial_vd-\partial_vc\,\partial_zd\right)\notag+z\left(\partial_uc\,\partial_vd-\partial_vc\,
\partial_ud\right)\notag \\
&&+\partial_z(cd)-c\,\partial_vc-d\,\partial_ud=\alpha^2\,,\label{s2}\\
&&ad-bc=1\,,\label{s3}
\end{eqnarray}
with additional condition $a=1+\dots$, $d=1+\dots$ reproducing the  commutative/flat regime.

The expression \eqref{pbxp} determines $\gamma_{\mu\nu}(x,p)$ once $a,b,c,d$ are fixed. The system \eqref{s1}–\eqref{s3} is the basis for all the realizations of our interest. In the following subsections we will analyze three particularly simple and useful families: Ansatz v-dependent, Ansatz u-dependente and Ansatz z-dependent.
In the next subsection, we start with the v-dependent case and explicitly show the solutions and the form of $\gamma_{\mu\nu}(x,p)$.

\subsection{$z$-dependent solution.}
Let us look for the solution of the system (\ref{s1}-\ref{s3}) in the assumption that all functions $a$, $b$, $c$ and $d$ may depend only on $z$-variable. Taking into account that $\{a(z),b(z)\}=0$ and consequently $\partial_u = \partial_v=0$, so our system will reduce to
\begin{equation}
\partial_z(ab)=\beta^2\,,\qquad \partial_z(cd)=\alpha^2\,,\qquad ad-bc=1\,.
\end{equation}
From the first and second equations one arrives at 
\begin{equation}\label{M6}
b=\frac{\beta^2z+k_5}{a}\,,\qquad c=\frac{\alpha^2z+k_6}{d}\,.
\end{equation}
Then the third equation gives,
\begin{equation}
(ad)^2-ad-\left(\beta^2z+k_5\right)\left(\alpha^2z+k_6\right)=0\,.
\end{equation}
It is a quadratic equation with the solution which satisfies our requirement given by,
\begin{equation}\label{ad}
ad=\frac{1\pm\sqrt{1+4\left(\beta^2z+k_5\right)\left(\alpha^2z+k_6\right)}}{2}\,.
\end{equation}
Starting from here different possibilities can be considered for the constants $k_5$ and $k_6$ aiming simplification of (\ref{ad}) and the functions $a$, $b$, $c$ and $d$ correspondingly. The most simple situation is the case of linear functions. And for this the argument of the square root in (\ref{ad}) should become a square, $(\dots)^2$. Taking into account that the functions $b$ and $c$ given by (\ref{M6}) also should be linear, the only possibility is,
\begin{equation*}
k_5=\pm\frac{\beta}{\alpha}\left(t-\frac12\right)\qquad\mbox{and}\qquad k_6= \pm\frac{\alpha}{\beta}\left(t+\frac12\right)
\end{equation*}
yielding, 
\begin{equation*}
ad=\frac12\pm\left(\alpha\beta z\pm t\right).
\end{equation*}
The correct commutative and ``flat'' limits require, $\lim_{\alpha\to0;\beta\to0}ad=1$. So, one takes $t=\pm 1/2$.

\paragraph{Possibility 1.} We start setting $k_5=0$ and $k_6=-\alpha/\beta$. In this case, taking $a=1$ one finds,
\begin{equation}\label{e1}
d=1-\alpha\,\beta\,z\,,\qquad b=\beta^2z\,,\qquad c=-\frac{\alpha}{\beta}\,.
\end{equation}
Which yields the linear realization (cf. eq. (14) in  \cite{mignemi_quantum_2016} considering $\lambda= 1-\alpha \beta \bar{z}$),
\begin{eqnarray}\label{e2}
x^\mu&=&\bar y^\mu+\beta^2\bar z\bar\xi^\mu\,,\\
p_\mu&=&-\frac{\alpha}{\beta}\,\bar y_\mu+\left(1-\alpha\,\beta\,\bar z\right)\bar\xi_\mu\,,\notag
\end{eqnarray}
with the inverse transformation given by,
\begin{eqnarray}\label{e3}
\bar y^\mu&=&x^\mu-\frac{\alpha\,x^2+\beta(x\cdot p)}{1+(\alpha x+\beta p)^2}\left(\alpha\,x^\mu+\beta\,p^\mu\right)\,,\\
\bar\xi^\mu&=&p^\mu+\frac{\alpha}{\beta}\,x^\mu\,,\notag
\end{eqnarray}
Using (\ref{e1}), (\ref{e2}) and (\ref{e3}) in (\ref{pbxp}) one finds,
\begin{eqnarray}
\label{e4}
\{x^\mu,p^\nu\}&=&\eta^{\mu\nu}-\alpha\beta\,\bar y^\mu\,\bar\xi^\nu+\alpha\beta\,\bar\xi^\mu\,\bar y^\nu+\beta^2\bar\xi^\mu\,\bar\xi^\nu\\
&=&\eta^{\mu\nu}+\alpha^2x^\mu\,x^\nu+\beta^2p^\mu\,p^\nu+2\alpha\beta\,x^\nu\,p^\mu\,,\notag
\end{eqnarray}
which is exactly the case of Snyder-de-Sitter algebra \cite{kowalski-glikman_triply_2004} in the semi-classical approximation. Here we also note that the representation (\ref{e2}) does not admit well defined commutative limit, $\beta\to0$.

\paragraph{Possibility 2.} Another possibility is to choose $k_5=-\beta/\alpha$, $k_6=0$ and $d=1$ leading to 
\begin{eqnarray}\label{e5}
x^\mu&=&\left(1-\alpha\,\beta\,\tilde z\right)\tilde y^\mu-\frac{\beta}{\alpha}\,\tilde\xi^\mu\,,\\
p^\mu&=&\alpha^2\tilde z\,y^\mu+\tilde\xi^\mu\,.\notag
\end{eqnarray}
In this case the ``flat'' limit $\alpha\to0$ is not well defined. 
The inverse transformation becomes,
\begin{eqnarray}\label{e6}
\tilde y^\mu&=&x^\mu+\frac{\beta}{\alpha}\,p^\mu\,,\\
\tilde\xi^\mu&=&p^\mu-\frac{\alpha(x\cdot p)+\beta\,p^2}{1+(\alpha x+\beta p)^2}\left(\alpha\,x^\mu+\beta\,p^\mu\right)\,.\notag
\end{eqnarray}
Again, using (\ref{e5}) and (\ref{e6}) in (\ref{pbxp}) one recovers the Poisson bracket (\ref{e4}) corresponding to the Snyder-de-Sitter case. 

\paragraph{Possibility 3.}  Any other choice of the constants $k_5$ and $k_6$ will yield nonlinear in $z$ functions $a$, $b$, $c$ and $z$. So, the corresponding bracket $\{x^\mu,p^\nu\}$ will be different from (\ref{e4}) and the corresponding algebra will not be Snyder-de-Sitter \cite{kowalski-glikman_triply_2004}. Just as a matter of illustration one may take $k_5=k_6=0$. In such case, taking 
\begin{eqnarray}\label{M8}
a&=&1\,, \qquad b=\beta^2z\,,\\
d&=&\frac{1+\sqrt{1+4\alpha^2\beta^2z^2}}{2}\,,\qquad c=\frac{2\,\alpha^2\,z}{1+\sqrt{1+4\alpha^2\beta^2z^2}}\,
\end{eqnarray}
we arrive at
\begin{eqnarray}\label{e7}
x^\mu=\bar y^\mu+\beta^2z\, \bar \xi^\mu\,,\qquad p_\mu=\frac{2\,\alpha^2\,z}{1+\sqrt{1+4\alpha^2\beta^2z^2}}\,\bar y_\mu+\frac{1+\sqrt{1+4\alpha^2\beta^2z^2}}{2}\, \bar \xi_\mu\,.
\end{eqnarray}
which is free of ratios $\alpha/\beta$ or $\beta/\alpha$ and admits both the flat $(\alpha \to 0)$ and commutative $(\beta \to 0)$ limits. To write the inverse transformation first we need to express $z$ in terms of $x^2$, $p^2$ and $(x\cdot p)$. From (\ref{M4}) one finds,
\begin{equation}
z=\left(ad+bc\right) (x\cdot p)-ab\,p^2-cd\,x^2\,.
\end{equation}
Now using (\ref{M6}) and (\ref{s3}) we rewrite it as,
\begin{equation}
z=\left(2\,ad-1\right) (x\cdot p)-\beta^2z\,p^2-\alpha^2z\,x^2\,,
\end{equation}
or
\begin{equation}
\left(1+\alpha^2\,x^2+\beta^2\,p^2\right)z=\sqrt{1+4\alpha^2\beta^2z^2}\, (x\cdot p)\,,
\end{equation}
implying that,
\begin{equation}\label{z}
z=\left(m^2-4\,\alpha^2\beta^2\right)^{-\frac12}, \qquad \mbox{with}\qquad m=\frac{1+\alpha^2x^2+\beta^2p^2}{(x\cdot p)}\,.
\end{equation}
In this case,
\begin{eqnarray}
d=\frac{\sqrt{m^2-4\,\alpha^2\beta^2}+m}{2\sqrt{m^2-4\,\alpha^2\beta^2}}\qquad\mbox{and}\qquad c=\frac{2\,\alpha^2}{\sqrt{m^2-4\,\alpha^2\beta^2}+m}\,.
\end{eqnarray}

Using (\ref{pbxp}) and the expression for $\bar y_\mu$ and $\bar\xi_\mu$ in terms of $x$ and $p$ we calculate the Poisson bracket,
\begin{eqnarray}
\{x_\mu,p_\nu\}= \eta_{\mu\nu}+k_1(m)\,x_\mu\,x_\nu+k_2(m)\,p_\mu\,p_\nu+k_3(m)\,\,x_\nu \,p_\mu\,,
\end{eqnarray}
where 
\begin{eqnarray}
k_1(m)&=&\frac{\alpha^2}{2}\,\frac{m+\sqrt{m^2-4\alpha^2\beta^2}}{\sqrt{m^2-4\alpha^2\beta^2}}\,,\\
k_2(m)&=&\beta^2\,,\\
k_3(m)&=&-\frac{4\alpha^2\beta^2}{m}\,,
\end{eqnarray}
and $m$ is given by (\ref{z}). So, as it has been already mentioned we are not getting here exactly the Snyder-de-Sitter model (\ref{e4}). However, the representation (\ref{e7}) admits both ``flat'' $\alpha\to0$ and commutative $\beta\to0$ limits.

\subsection{$v$-dependent solution}
Now let us suppose that all functions may depend only on $v$ variable. Then, the system (\ref{s1}-\ref{s3})  becomes,
\begin{equation}
-a\,\partial_va=\beta^2\,,\qquad -c\,\partial_vc=\alpha^2\,,\qquad ad-bc=1\,.
\end{equation}
From the first two equations one finds,
\begin{equation}
a=\sqrt{k_1-2\beta^2v}\qquad \mbox{and}\qquad c=\pm\sqrt{k_2-2\alpha^2v}\,.
\end{equation}
The boundary condition $a=1+\dots$ implies that $k_1=1$, and for simplicity taking, $k_2=\alpha^2/\beta^2$, we arrive at
\begin{equation}
    a= \sqrt{1-2\be^2 v}, \qquad c= \pm \frac{\a}{\be}\sqrt{1-2\be^2 v}.
\end{equation}
The algebraic constraint $ad-bc=1$ is solved by
\begin{equation}
d=\frac{1}{a} \qquad\mbox{and}\qquad b=0\,,
\end{equation}
and, choosing the minus sign for $c$, one obtains the realization
\begin{eqnarray}\label{e8}
x_\mu&=&\bar y_\mu\sqrt{1-\beta^2\,\bar\xi^2}\,, \\
p_\mu&=&-\frac{\alpha}{\beta}\, \bar y_\mu\sqrt{1-\beta^2\,\bar \xi^2} +\frac{\bar \xi_\mu}{ \sqrt{1-\beta^2\,\bar \xi^2}}\,.\notag
\end{eqnarray}

Just like in the case of (\ref{e2}) this realization does not admit a well defined commutative limit, since $\beta$ appears in the denominator.  
The inverse transformation reads,
\begin{eqnarray}\label{inversee8}
\bar y_\mu&=&x_\mu\,\sqrt{1+(\beta p+\alpha x)^2}\,,\\
\bar \xi_\mu&=&\frac{1}{\beta}\,\frac{\beta\,p_\mu+\alpha\,x_\mu}{\sqrt{1+(\beta p+\alpha x)^2}}\,.\notag
\end{eqnarray}
Using (\ref{pbxp}) one calculates,
\begin{eqnarray}
\{x_\mu,p_\nu\}&=&\eta_{\mu\nu}+\alpha\beta\left(\bar y_\mu\,\bar \xi_\nu-\bar \xi_\mu\,\bar y_\nu\right)+\frac{\beta^2\,\bar \xi_\mu\,\bar \xi_\nu}{1-\beta^2\,\bar \xi^2} \\
&=&\eta_{\mu\nu}+\alpha\,\beta\left(x_\mu\,p_\nu-x_\nu\,p_\mu\right)+\beta^2\left(p_\mu+\frac{\alpha}{\beta}\,x_\mu\right)\left(p_\nu+\frac{\alpha}{\beta}\,x_\nu\right)\notag\\
&=&\eta_{\mu\nu}+\alpha^2\,x_\mu\,\,x_\nu+\beta^2\,p_\mu\,p_\nu+2\alpha\beta\,x_\mu \,p_\nu\,.\notag
\end{eqnarray}
which is exactly the Snyder–de Sitter mixed bracket. As in the z-linear realization \eqref{e2}, the presence of the ratio $\alpha/\beta$ makes the commutative limit $\beta \to 0$ non-uniform (unless $\alpha/\beta$ is scaled simultaneously).

\subsection{$u$-dependent solution}

Finally let us consider the situation when all functions may depend only on $u$ variable. Then, the system (\ref{s1}-\ref{s3})  becomes,
\begin{equation}
-b\,\partial_ub=\beta^2\,,\qquad -d\,\partial_ud=\alpha^2\,,\qquad ad-bc=1\,.
\end{equation}
From the first two equations one finds,
\begin{equation}
b=\pm\sqrt{k_3-2\beta^2u}\qquad \mbox{and}\qquad d=\sqrt{k_4-2\alpha^2u}\,.
\end{equation}
The condition $d=1+\dots$ implies that $k_4=1$, and for simplicity we take, $k_3=\beta^2/\alpha^2$, $c=0$ and $a=1/d$. With the minus sign for $b$ on arrives at,
\begin{eqnarray} \label{e9}
x_\mu&=&\frac{\tilde y_\mu}{ \sqrt{1-\alpha^2\,\tilde y^2}}-\frac{\beta}{\alpha} \tilde \xi_\mu\sqrt{1-\alpha^2\,\tilde y^2} \,,\\
p_\mu&=&\tilde \xi_\mu\sqrt{1-\alpha^2\,\tilde y^2}\,.\notag
\end{eqnarray}
This is the u-dual of the v-dependent realization.

As in the the v-dependent case, a ratio os scales appears ($\beta/\alpha$, here), so the flat limit $\alpha \to 0$ is not uniform well defined unless the ratio is scaled appropriately. The inverse is given by
\begin{eqnarray}
\tilde{y}_\mu&=& \frac{\alpha x_\mu\ + \beta p_\mu}{\alpha \sqrt{1+(\beta p+\alpha x)^2}} \,,\\
\tilde{\xi}_\mu&=&p_\mu \sqrt{1+(\beta p+\alpha x)^2}.\notag
\end{eqnarray}
Again using (\ref{pbxp}) one may check that the constructed representation (\ref{e9}) closes the SdS algebra (\ref{SdS}).

\subsection{Arbitrariness and canonical transformations}

The equations (\ref{e2}), (\ref{e5}), (\ref{e8}) and (\ref{e9}) represent the same SdS-algebra (\ref{SdS}) in terms of different sets of the canonical variables, so there should exist a canonical transformation connecting them. In this subsection we construct an explicit form of these canonical transformations and make a comment regarding the arbitrariness in our construction. We start constructing a canonical transformation between variables $(\bar y^\mu, \bar\xi^\nu)$ and $(\tilde y^\mu, \tilde\xi^\nu)$ entering (\ref{e2}) and (\ref{e5}) correspondingly. An explicit (and convenient) way to display it is as the composition
\begin{equation*}
    (\bar{y}, \bar{\xi})
\xrightarrow{\eqref{e2}}
(x, p)
\xrightarrow{\eqref{e6}}
    (\tilde{y}, \tilde{\xi})
\end{equation*}
which is canonical by construction,
\begin{equation}\label{t1}
    \tilde{y}^\mu = \frac{\beta}{\alpha} \, \xi^\mu, \qquad
    \tilde{\xi}^\mu = -\frac{\alpha}{\beta} \, \bar{y}^\mu + \frac{1}{1 + \beta^2 \bar{\xi}^2} \, \xi^\mu \qquad \text{with}\qquad \bar{\xi}^2 := \eta_{\rho \sigma} \, \bar{\xi}^\rho \bar{\xi}^\sigma.
\end{equation}
For later use, note the identity
\begin{equation*}
    \a x+\be p= \be \bar{\xi}, 
\end{equation*}
from which one immediately gets $\tilde{y}= x+\frac{\be}{\a}\bar{\xi}$; the corresponding $\tilde{\xi}$ follows from \eqref{e6}.

Now let us derive canonical transformation relating \eqref{e2} and \eqref{e8}.
From the z-linear realization \eqref{e2} one has the identity
\begin{equation}
    1+ (\beta p + \alpha x)^2 = 1+ (\beta \bar{\xi})^2.
\end{equation}
Substituting this in \eqref{inversee8} gives immediately
\begin{equation}\label{t2}
    \xi_{\mu}= \frac{\bar{\xi}_{\mu}}{\sqrt{1+\beta^2 \bar{\xi}^2}}, \qquad y_\mu = (\bar{y}_{\mu}+ \beta^2 \bar{z}\bar{\xi}_\mu) \sqrt{1+\beta^2 \bar{\xi}^2}, 
\end{equation}
with $\bar{z}= \bar{y}\cdot\bar{\xi}$. This map carries the Darboux pair $(\bar{y}, \bar{\xi})$ of \eqref{e2} to the Darboux pair $(y, \xi)$ of \eqref{e8} and is canonical.
A convenient way to exhibit canonicity is to use the generating function of type 2,
\begin{equation}
    \mathcal{F}_2^{(\beta, \Lambda)}(\bar{y}, \tilde{\xi}) = \frac{(\Lambda \bar{y}) \cdot \tilde{\xi}}{\sqrt{1 - \beta^2 \tilde{\xi}^2}}, \qquad \Lambda^T \Lambda = \mathbb{I}.
\end{equation}
Then,
\begin{equation}
    \bar{\xi} = \frac{\partial \mathcal{F}_2}{\partial \bar{y}} = \Lambda^T \frac{\tilde{\xi}}{\sqrt{1 - \beta^2 \tilde{\xi}^2}} 
    \quad \Rightarrow \quad 
    \tilde{\xi} = \frac{\Lambda \bar{\xi}}{\sqrt{1 + \beta^2 \bar{\xi}^2}},
\end{equation}
and
\begin{equation}
    \tilde{y} = \frac{\partial \mathcal{F}_2}{\partial \tilde{\xi}} = \frac{\Lambda \bar{y}}{\sqrt{1 - \beta^2 \tilde{\xi}^2}} 
    + \frac{\beta^2 [(\Lambda \bar{y}) \tilde{\xi})]}{(1 - \beta^2 \tilde{\xi}^2)^{3/2}} \tilde{\xi}
    \quad \Rightarrow \quad 
    \tilde{y} = \Lambda \left( \bar{y} + \beta^2 \bar{z} \bar{\xi} \right) \sqrt{1 + \beta^2 \bar{\xi}^2}.
\end{equation}
Setting $\Lambda=\id$ recovers the explicit formulas above. 

Canonical transformation relating  $v$-dependent realization \eqref{e8} and $u$-dependent one \eqref{e9} is given by the formula,
\begin{equation}\label{t3}
    \tilde y_\mu=\frac{\beta}{\alpha}\,\xi_\mu,\qquad
\tilde\xi_\mu=\frac{\xi_\mu}{1-\beta^2\xi^2}-\frac{\alpha}{\beta}\,y_\mu\,.
\end{equation}
Thus, the composition of the transformations (\ref{t1}), (\ref{t2}), and (\ref{t3}), together with their respective inverses, establishes the correspondence among all canonical representations of the SdS-algebra (\ref{SdS}) considered in this section. It should be emphasized that, in the present analysis, we have restricted attention to the most elementary solutions of the system (\ref{s1}–\ref{s3}). Any alternative canonical phase-space representation of (\ref{SdS}) can be derived from those constructed here by means of an appropriate canonical transformation, thereby reflecting the intrinsic arbitrariness underlying our construction.

\subsection{Yang model.} 

Taking in the system  \eqref{s1}–\eqref{s3}, $b=c=0$, one it reduces to,
\begin{equation}
\partial_va^2=-2\beta^2\,,\qquad \partial_ud^2=-2\alpha^2\,,\qquad ad=1\,.
\end{equation}
The first two equations imply,
\begin{equation}
a^2=k_1(u,z)-2\beta^2v\,,\qquad d^2=k_2(v,z)-2\alpha^2u\,,
\end{equation}
with $k_1(u,z)$ and $k_2(v,z)$ being arbitrary functions of indicated arguments which should satisfy
\begin{equation}\label{M7}
k_1(u,z)k_2(v,z)-2\alpha^2u\,k_1(u,z)-2\beta^2v\,k_2(u,z)+4\alpha^2\beta^2\,u\,v=1\,.
\end{equation}
This functional equation in very restrictive; under condition $ad=1$ it seems not to admit nontrivial solutions with $k_1, k_2$ depending on $(u,z),(v,z)$ independently.

However, the authors of \cite{Meljanac:2023dpr} relax the condition (\ref{s3}), which implies that (\ref{M7}) is no longer required. In this case, 
\begin{equation}
M_{\mu\nu}=\frac{x_\mu\,p_\nu-x_\nu\,p_\mu}{ad}\,.
\end{equation}
If in addition, $\partial_ua=\partial_vd=0$, i.e., $k_1=k_1(z)$ and $k_2=k_2(z)$, and also
\begin{equation}\label{q1}
\{a,d\}+\partial_z(ad)=0\,,
\end{equation}
then by (\ref{pbxp}) one finds,
\begin{equation}
\{x_\mu,p_\nu\}=\eta_{\mu\nu}\,ad\,.
\end{equation}
The equation (\ref{q1}) implies,
\begin{equation}
k_1(z)k_2(z)=\alpha^2\beta^2z^2+const.
\end{equation}
The $const$ should be taken to be $1$ to guarantee the correct commutative and flat limits. Then, one may set,
\begin{equation}
k_1(z)=k_2(z)=\sqrt{1+\alpha^2\beta^2z^2}\,,
\end{equation}
providing the Born reciprocity. In doing so one finds,
\begin{eqnarray}
x_\mu=\bar y_\mu\,\sqrt{\sqrt{1+\alpha^2\beta^2z^2}-\beta^2\bar\xi^2}\,,\qquad p_\mu=\bar\xi_\mu ,\sqrt{\sqrt{1+\alpha^2\beta^2z^2}-\alpha^2\bar y^2}\,,
\end{eqnarray}
which is in agreement with \cite{Meljanac:2023dpr} and provides the representation of the Yang model (\ref{h}) with
\begin{equation}
h=ad=\sqrt{1-\alpha^2x^2-\beta^2p^2-\frac{\alpha^2\beta^2}{2}\,M^2}\,.
\end{equation}
 At the same time it does not reproduces neither Snyder (\ref{Sn}) in flat limit $\alpha\to0$, nor de-Sitter algebras in the commutative limit $\beta\to0$.

\section{Flat and commutative limits and geometric interpretation}

To gain a clearer understanding of the phase-space geometry of the model under consideration, we analyze in this section two distinct regimes: (i) the flat limit $\alpha \to 0$ (corresponding to the curvature scale $R \to \infty$), and (ii) the commutative limit $\beta \to 0$ (corresponding to vanishing coordinate noncommutativity). The behavior of different Darboux charts is not uniform under these limits; therefore, we explicitly indicate which realizations exhibit consistent behavior.

\paragraph{Flat limit.} We start with the limit $\a \to 0$ considering the original Snyder algebra (\ref{Sn}). It is known, see e.g. \cite{banerjee2006deformed,Mignemi:2011gr}, that the momenta space of this model is curved. More precisely, it is a four dimensional hyperboloid embedded in five dimensional Minkowski space through the equation, 
\begin{equation}\label{dSp}
\bar\xi_A \bar\xi^A := -\bar\xi_0^2 + \bar\xi_1^2 + \bar\xi_2^2 + \bar\xi_3^2 + \bar\xi_4^2=\frac{1}{\beta^2}\,,
\end{equation}
which is equivalent to
\begin{equation}\label{dSp1}
\beta\bar\xi_4=\pm\sqrt{1-\beta^2\bar\xi^2}\,\qquad \bar\xi^2=\bar\xi_\mu\bar\xi^\mu\,.
\end{equation}
The corresponding action reads
\begin{equation}\label{aSn}
S_{Snyder}=\int d\tau\left[\bar\xi_A\,\dot{\bar y}^A-H(\bar y,\bar \xi)+\lambda\left(\bar\xi_A\bar\xi^A-\beta^{-2}\right)\right],
\end{equation}
where condition (\ref{dSp}) was introduced as a Hamiltonian constraint with the Lagrangian multiplier $\lambda$. The phase space variables $\bar y^A$ and $\bar\xi_A$ are standard canonical variables satisfying (\ref{Dc2na})  and $H(\bar y,\bar \xi)$ is some given Hamiltonian describing the dynamics of the system which preserves in time the constraint (\ref{dSp}). Now let us introduce the Beltrami coordinates on the hyperboloid (\ref{dSp}),
\begin{equation}\label{pSn}
p_\mu=\frac{\bar\xi_\mu}{\beta \bar\xi_4}=\frac{\bar\xi_\mu}{\sqrt{1-\beta^2\bar\xi^2}}\,,
\end{equation}
which are declared to be physical momenta. And physical coordinates are expressed in terms of original canonical variables as,
\begin{equation}\label{xSn}
x_\mu=\bar y_\mu\,\beta \bar\xi_4=\bar y_\mu\,\sqrt{1-\beta^2\bar\xi^2}\,.
\end{equation}
The expressions (\ref{pSn}) and (\ref{xSn}) are exactly the same as (\ref{e8}) with $\alpha=0$. Using the expression for the inverses,
\begin{equation}
\bar\xi_\mu=\frac{p_\mu}{\sqrt{1+\beta^2p^2}}\,,\qquad \beta\bar\xi_4=\frac{1}{\sqrt{1+\beta^2p^2}}\,,\qquad \bar y^\mu=x^\mu\sqrt{1+\beta^2p^2}\,,
\end{equation}
one rewrites the action (\ref{aSn}) in terms of physical phase space variables as follows,
\begin{equation}\label{aSn1}
S_{Snyder}=\int d\tau\left[p_\mu\dot x^\mu+\frac{(p\cdot x)p_\mu\dot p^\mu}{1+\beta^2p^2}-H\right].
\end{equation}
The corresponding Poisson structure is exactly Snyder algebra (\ref{Sn}). Choosing the free particle Hamiltonian as $H=p^2/2$ one obtains the free fall equations,
\begin{equation}
\dot x^\mu=\left(1+\beta^2\,p^2\right)p^\mu\,,\qquad \dot p_\mu=0\,,
\end{equation}
with the solution \cite{Mignemi:2011gr}, $p_\mu=const$ and $x^\mu=\dot x_0^\mu\tau+ x_0^\mu$. Although the momentum space of the system is curved, the free-fall trajectories remain straight lines, as in the commutative case. Finally, it should be noted that a curved momentum space is a characteristic feature of Poisson electrodynamics \cite{Kupriyanov:2024dny,Basilio:2024bir,Sharapov:2025lik}, with the precise geometric structure determined by the specific choice of coordinate noncommutativity.

\paragraph{Commutative limit.} The very similar situation happens in the limit $\beta\to0$ with the difference that now the momenta space is flat whilst the configuration space is hyperboloid defined by the equation, $\bar y_A \bar y^A=1/\alpha^2$. And the action in the embedding space reads,
\begin{equation}\label{adS}
S_{dS}=\int d\tau\left[\bar\xi_A\,\dot{\bar y}^A-H(\bar y,\bar \xi)+\lambda\left(\bar y_A\bar y^A-\alpha^{-2}\right)\right].
\end{equation}
In this case the Beltrami coordinates on the hyperboloid,
\begin{equation}
x^\mu=\frac{\bar y^\mu}{\alpha\bar y_4}=\frac{\bar y^\mu}{\sqrt{1-\alpha^2\bar y^2}}\,,
\end{equation}
are denominated as physical coordinates with the physical momenta being,
\begin{equation}
p_\mu=\sqrt{1-\alpha^2\bar y^2}\,\bar\xi_\mu\,.
\end{equation}
The latter definitions coincide explicitly with (\ref{e9}) for $\beta=0$. The action (\ref{adS}) in new coordinates becomes,
\begin{equation}\label{Haction}
S_{dS}=\int d\tau\left[\left(p_\mu-\frac{\alpha^2\left(x\cdot p\right)x_\mu}{1+\alpha^2x^2}\right)\dot x^\mu-H\right],
\end{equation}
reproducing de-Sitter algebra for corresponding Poisson brackets,
\begin{equation}\label{dSa1}
\{x_\mu,x_\nu\}=0\,,\qquad \{x_\mu,p_\nu\}=\eta_{\mu\nu}+\alpha^2x_\mu x_\nu\,,\qquad \{p_\mu,p_\nu\}=\alpha^2\left(x_\mu p_\nu-x_\nu p_\mu\right).
\end{equation}
The corresponding free particle trajectories are geodesics on de-Sitter space, see \cite{Mignemi:2008fj} for more details.

\section{Symplectic realizations and deformed partial derivative}  

In this final section, we return to the problem posed in the introduction: the construction of a symplectic realization of the generalized Snyder–Poisson algebra and the definition of a deformed partial derivative that differentiates the corresponding Poisson brackets. The method we develop is applicable to any generalization of the Snyder–Poisson algebra (\ref{Sn}) or the de Sitter algebra (\ref{dSa1}). For the sake of clarity and concreteness, we focus here on the Snyder–de Sitter algebra (\ref{SdS}). To proceed, we first introduce a more convenient notation for this algebra,
\begin{eqnarray}
\label{Sn1}
&& \{X^{\cal M},X^{\cal N}\}=\Theta^{{\cal M}{\cal N}}_{SdS}(X)\,, \\ \notag
&& \Theta^{{\cal M}{\cal N}}_{SdS}=\left({\begin{array}{cc}
 \beta^2\left(x_\mu\,p_\nu-x_\nu\,p_\mu\right) & \eta_{\mu \nu}+\alpha^2 x_\mu x_\nu+\beta^2 p_\mu p_\nu+2\alpha\beta p_\mu x_\nu\\
 -\eta_{\mu \nu}-\alpha^2 x_\nu x_\mu-\beta^2 p_\nu p_\mu-2\alpha\beta p_\nu x_\mu
 & \alpha^2\left(x_\mu p_\nu-x_\nu p_\mu\right)
 \end{array}}\right),
\end{eqnarray}
with  $X^{\mathcal{M}}= (x_\mu, p_\mu)$. 
The principal technical challenge in constructing a Poisson gauge theory on the specified noncommutative space arises from the fact that the standard partial derivative $\partial_{\cal M} := \partial / \partial X^{\cal M}$ does not act as a derivation of the Poisson structure (\ref{Sn1}). In particular, the Leibniz rule fails to hold, $\partial_{\cal M}\{f,g\}\neq \{\partial_{\cal M}f,g\}+\{f,\partial_{\cal M}g\}$.  To address this difficulty, we first construct a symplectic realization of (\ref{Sn1}), and subsequently introduce a deformed, or “twisted,” partial derivative  $\bar\partial_{\cal M}$ that restores the desired Leibniz property.

\paragraph{Symplectic realization of the algebra (\ref{Sn1}).} Essentially, to construct the symplectic realization of the algebra (\ref{Sn1}) which is suitable for the Poisson gauge theory one needs to introduce additional $2N$ variables $\xi_{\cal M}=(\xi_\mu,\xi_\ast^\mu)$, in such a way that the Poisson brackets,
\begin{equation}\label{srSn1}
\{X^{\cal M},X^{\cal N}\}=\Theta^{{\cal M}{\cal N}}(X)\,,\qquad\{X^{\cal M},\xi_{\cal N}\}=\Gamma^{{\cal M}}_{{\cal N}}(X,\xi)\,, \qquad \{\xi_{\cal M},\xi_{\cal N}\}=0\,,
\end{equation}
satisfy the Jacobi identity. For convenience, we impose that the Poisson brackets between the auxiliary variables $\xi_{\cal M}$ vanish. The antisymmetric tensor $\Theta^{{\cal M}{\cal N}}(X)$ is taken as given. In this setting, constructing a symplectic realization amounts to determining the matrix $\Gamma^{{\cal M}}_{{\cal N}}(X,\xi)$ in such a way that the Jacobi identity for (\ref{srSn1}) holds true.

Since the proper algebra (\ref{Sn1}) is already the symplectic one, to construct its symplectic realization we will use the Darboux (canonical) coordinates $Y^{\cal M}=(y^\mu,y^\ast_\mu)$ and $\xi_{\cal N}=(\xi_\nu,\xi_\ast^\nu)$ in $4N$-dimmensional space, satisfying the commutation relations,
\begin{equation}\label{Dc4n}
 \{Y^{\cal M},Y^{\cal N}\}=0\,,\qquad\{Y^{\cal M},\xi_{\cal N}\}=\delta^{\cal M}_{\cal N},\qquad \{\xi_{\cal M},\xi_{\cal N}\}=0\,.
\end{equation}
Now let us introduce,
\begin{equation}\label{Dc2n}
\bar y^\mu=y^\mu-\frac12\,\xi_\ast^\mu\qquad\mbox{and}\qquad \bar\xi_\mu=y^\ast_\mu+\frac12\,\xi_\mu\,,
\end{equation}
satisfying (\ref{Dc2na}).
In fact, they are also Darboux coordinates, however in $2n$-dimensional  space. Different possibilities to represent the original phase space coordinates $x^\mu$ and $p_\mu$ in terms of the Darboux coordinates (\ref{Dc2n}) were discussed in Section 2. One may choose, for example, the expression for $X^{\cal M}(\bar y,\bar\xi)$ given by (\ref{e8}). Then,
\begin{equation}\label{defG}
\Gamma^{{\cal M}}_{{\cal N}}(\bar y,\bar\xi):=\{X^{\cal M}(Y,\xi),\xi_{\cal N}\}=\frac{\partial X^{\cal M}}{\partial Y^{\cal L}}\{Y^{\cal L},\xi_{\cal N}\}=\frac{\partial X^{\cal M}}{\partial Y^{\cal N}}\,,
\end{equation}
and applying the inverse given by (\ref{inversee8}) one finds $\Gamma^{{\cal M}}_{{\cal N}}(X)$.  We write it carefully, 
\begin{eqnarray}
\{x^\mu,\xi_\nu\}&=&\delta^\mu_\nu\,\sqrt{1-\beta^2\,\bar\xi^2}=\frac{\delta^\mu_\nu}{\sqrt{1+(\alpha x+\beta p)^2}}\,,\\
\{x^\mu,\xi_\ast^\nu\}&=&\frac{-\beta^2\bar y^\mu\,\bar\xi_\nu}{\sqrt{1-\beta^2\,\bar\xi^2}}=-\beta x^\mu\left(\alpha x_\nu+\beta p_\nu\right)\sqrt{1+(\alpha x+\beta p)^2}\,,\notag\\
\{p^\mu,\xi_\nu\}&=&-\frac{\alpha}{\beta}\delta^\mu_\nu\,\sqrt{1-\beta^2\,\bar\xi^2}=\frac{-\alpha\,\delta^\mu_\nu}{\beta\sqrt{1+(\alpha x+\beta p)^2}}\,,\notag\\
 \{p_\mu,\xi_\ast^\nu\}&=&\frac{\left(\delta_\mu^\nu+\alpha\beta\,\bar y_\mu\,\bar\xi^\nu\right)(1-\beta^2\bar\xi^2)+\beta^2\bar\xi_\mu\bar\xi^\nu}{(1-\beta^2\bar\xi^2)^{3/2}} \notag\\
 &=&\left(\delta_\mu^\nu+\left(2\alpha x_\mu+\beta p_\mu\right)\left(\alpha x_\nu+\beta p_\nu\right)\right)\sqrt{1+(\alpha x+\beta p)^2}\,.\notag
\end{eqnarray}
Which yields,
\begin{eqnarray}\label{Gamma}
\Gamma^{{\cal M}}_{{\cal N}}(X)=\left({\begin{array}{cc}
\frac{\delta^\mu_\nu}{\sqrt{1+(\alpha x+\beta p)^2}} & -\beta x^\mu\left(\alpha x_\nu+\beta p_\nu\right)\sqrt{1+(\alpha x+\beta p)^2}\\
\frac{-\alpha\,\delta^\mu_\nu}{\beta\sqrt{1+(\alpha x+\beta p)^2}}
 &\left(\delta_\mu^\nu+\left(2\alpha x_\mu+\beta p_\mu\right)\left(\alpha x_\nu+\beta p_\nu\right)\right)\sqrt{1+(\alpha x+\beta p)^2}
 \end{array}}\right).
\end{eqnarray}
It is evident that the symplectic realization of a given Poisson structure is not unique \cite{PGT}. The realization defined by (\ref{Gamma}) does not admit a well-defined commutative limit $\beta \to 0$ for $\alpha$ fixed, as it is derived from the canonical representation (\ref{e8}) of the Snyder–de Sitter algebra (\ref{SdS}), which involves the ratio $\alpha / \beta$. Alternatively, one may employ, for instance, the representation (\ref{e9}), in which the commutative limit is well defined. However, in this case the flat limit $\alpha \to 0$ fails to exist, since the representation contains a term proportional to $\beta / \alpha$.

To conclude this part we note that the relation $\{X^{\cal M}(\bar y,\bar\xi),X^{\cal N}(\bar y,\bar\xi)\}=\Theta^{\cal MN}(X^{\cal M}(\bar y,\bar\xi))$ implies the following identity for the matrix $\Gamma^{{\cal M}}_{{\cal K}}(X)$ of the symplectic realization,
\begin{equation}\label{thetagamma}
\Gamma^{{\cal M}}_{{\cal K}}(X)\,\Theta_0^{{\cal K}{\cal L}}\,\Gamma^{{\cal N}}_{{\cal L}}(X)=\Theta^{{\cal M}{\cal N}}(X)\,,
\end{equation}
where $\Theta_0^{{\cal K}{\cal L}}$ is canonical symplectic matrix corresponding to the canonical Poisson structure (\ref{Dc2na}).

\paragraph{Deformed partial derivative.} Now let us return to the problem formulated in the beginning of this section, namely the fact that standard partial derivative $\partial_{\cal M}$ does not differentiate the Poisson structure (\ref{SdS}). We introduce the deformed partial derivative by the rule,
\begin{equation}\label{dpd}
\bar\partial_{\cal N}f(X):=\{f(X),\xi_{\cal N}\}=\Gamma^{\cal M}_{\cal N}(X)\,\partial_{\cal M}f(X)\,.
\end{equation}
In the limits, $\alpha\to0$ and $\beta\to0$, when $\Theta^{{\cal M}{\cal N}}(X)\to\Theta^{{\cal M}{\cal N}}_0$ and $\Gamma^{\cal M}_{\cal N}(X)\to\delta^{\cal M}_{\cal N}$, the deformed partial derivative (\ref{dpd}) becomes a standard partial derivative $\partial_{\cal N}$. However, the Jacoby identity for the algebra (\ref{srSn1}) implies the standard Leibniz rule for the Poisson bracket (\ref{Sn1}),
\begin{eqnarray}\label{dlr}
\bar\partial_{\cal N}\{f,g\}=\{\{f,g\},\xi_{\cal N}\}=\{\{f,\xi_{\cal N}\},g\}+\{f,\{g,\xi_{\cal N}\}\}=\{\bar\partial_{\cal N}f,g\}+\{f,\bar\partial_{\cal N}g\}\,,
\end{eqnarray}
which is one of main results of the current research.

\section{Conclusion}

To conclude this work, let us briefly outline potential applications of the proposed formalism. In the introduction, we formulated the problem of consistently defining gauge theory on the Snyder space (\ref{Sn}) and its generalizations, such as the Snyder–de Sitter space (\ref{SdS}). Having established the deformed partial derivative (\ref{dpd}), one can proceed to define the Poisson gauge transformation $\delta_f$ of the gauge field $A_{\cal M}(X)$. Following \cite{KS-21}, it takes the form
\begin{equation}\label{pgt}
\delta_fA_{\cal M}:=\{f(X), \xi_{\cal M}-A_{\cal M}(X)\}=\bar \partial_{\cal M} f+\{A_{\cal M},f\}\,.
\end{equation}
Using the deformed Leibniz rule (\ref{dlr}) one may easily check that Poisson gauge transformations close the algebra 
\begin{equation}
\left[\delta_f,\delta_g\right]=\delta_{\{f,g\}}\,.
\end{equation}
The field strength ${\cal F}_{\cal MN}(X)$ is defined according to \cite{PGT},
\begin{equation}\label{pfs}
{\cal F}_{\cal MN}(X):=\{\xi_{\cal M}-A_{\cal M}(X), \xi_{\cal N}-A_{\cal N}(X)\}=\bar \partial_{\cal M}A_{\cal N}-\bar \partial_{\cal N}A_{\cal M}+\{A_{\cal M},A_{\cal N}\}\,.
\end{equation}
Once again, employing the Leibniz rule (\ref{dlr}), one verifies that ${\cal F}_{\cal MN}$ transforms covariantly under the gauge transformation (\ref{pgt}), $\delta_f{\cal F}_{\cal MN}=\{f,{\cal F}_{\cal MN}\}$. The physical interpretation and implications of a gauge theory constructed on this basis will be addressed in forthcoming work. 

A further avenue of application lies in the deformation quantization of the associated Poisson structures, which naturally leads to the derivation of conserved currents of the form \begin{equation}
\bar \partial_{\cal M}j^{\cal M}(X)=0\,,
\end{equation}
 in the proposed formalism.

\subsection*{Acknowledgments}

We appreciate the support from the Funda\c c\~ao de Amparo a Pesquisa do Estado de S\~ao Paulo (FAPESP, Brazil), Grant numbers: $2024/04134-6$ and $2024/23831-0$. KVG also acknowledges the financial support from the Concelho Nacional de Pesquisa (CNPq, Brazil), grant number: $305132/2024-5$.

\end{document}